\documentclass[%
 reprint, 
superscriptaddress,
 amsmath,amssymb,
 aps, prl
]{revtex4-2}

\usepackage{graphicx}
\usepackage{amsmath}
\usepackage{mathrsfs}
\usepackage{setspace}
\usepackage{color}
\usepackage[dvipsnames,svgnames]{xcolor}
\usepackage{textcomp}
\usepackage{booktabs}
\usepackage[flushleft]{threeparttable}
\usepackage{comment}
\usepackage{amssymb}
\usepackage{amsmath}
\usepackage{float}
\usepackage{psfrag}
\usepackage{siunitx}
\usepackage{romannum}
\usepackage{bm}

\def\be{\begin{equation}}
\def\ee{\end{equation}}
\def\ba{\begin{eqnarray}}
\def\ea{\end{eqnarray}}

\usepackage{esdiff}

\begin{document}

\pagenumbering{arabic} 

% \preprint{APS/123-QED}

%\title{\textbf{Size reduction of bubble bursting jet drops due to a shallow liquid layer} 

\title{\normalsize\textbf{Confinement-Induced Suppression of Jet Drop Size by Bubble Bursting in Shallow Liquids} 
}% 

%\title{\textbf{Enhanced jet drop production by bubble bursting in a shallow liquid layer} 
%}% 

\author{Zhengyu Yang}
\affiliation{Mechanical Science and Engineering, University of Illinois Urbana-Champaign, Illinois 61801, USA}

\author{Vatsal Sanjay}
\affiliation{CoMPhy Lab, Department of Physics, Durham University,
Science Laboratories, South Road, Durham DH1 3LE, United Kingdom}

\author{C. Ricardo Constante-Amores}
\affiliation{Mechanical Science and Engineering, University of Illinois Urbana-Champaign, Illinois 61801, USA}

\author{Jie Feng}
\email{jiefeng@illinois.edu}
\affiliation{Mechanical Science and Engineering, University of Illinois Urbana-Champaign, Illinois 61801, USA}
\affiliation{Materials Research Laboratory, University of Illinois Urbana-Champaign, Illinois 61801, USA}


\begin{abstract} 
%We investigate the effect of a solid wall beneath a free bubble on the radius of aerosol drops produced by jetting during bubble bursting. The rupture of bubbles at a free surface disperses numerous jet drops into the air phase and plays a critical role in the aerosol generation in natural and industrial contexts. Although geometric confinements are common in realistic settings, their effects on jet drop formation are commonly neglected. 

Bubble bursting is a major source of aerosol generation in a wide range of natural and industrial systems. While the resulting jet dynamics have been extensively studied in deep liquid pools, bubble bursting often occurs in shallow liquid layers where the influence of the nearby solid boundary remains poorly understood. Here, we show numerically that a shallow liquid layer produces smaller and more numerous jet drops, even when the initial bubble shape is unchanged. We identify a wall-induced viscous sticking effect that suppresses the upward motion of the cavity bottom, leading to a steeper cavity geometry during capillary-wave focusing. We further develop a semi-empirical scaling law that predicts the jet drop radius as a function of the Ohnesorge number and the initial bubble-wall distance. Our results establish geometric confinement as a governing factor in bubble bursting and provide a framework for predicting and controlling aerosol generation in shallow liquid environments.

%Using numerical simulations, we observe that the presence of a nearby solid wall beneath the bubble leads to the production of smaller and more numerous jet drops even with the same bubble shape. We reveal that this behavior arises from a viscous sticking effect between the wall and the bubble. The viscous sticking effect produces a steeper cavity geometry when the capillary waves converge at the cavity bottom to generate the jet, and the effect together with the wave damping results in a reduced jet drop radius.

%Our work advances the fundamental understanding of bubble bursting and provides a framework for predicting similar free surface flows across different length scales. Furthermore, the study opens a new avenue for tuning the aerosol production from bubble bursting without altering liquid and bubble properties.
% TBD. (do mention the unexpected things. You can just start with "We investigate" without the long background/motivation explanation. Of course, "unexpected" statements will help.)
\end{abstract}

\maketitle

%\tableofcontents

The bursting of surface bubbles plays a critical role in multiple natural and industrial processes \citep{bird2010daughter,feng2014,veron2015ocean,deike2022mass,bourouiba2021fluid}. For example, drops ejected during bubble bursting are a primary source of sea spray aerosol generation and contribute significantly the ocean-atmosphere mass transfer \cite{deike2022mass}. Those bubble bursting drops has also been identified as an important pathway of contaminant and disease transmission needing control \cite{bourouiba2020,wang2017role}. Although most previous studies have focused on bubble bursting in deep liquid pools, bubbles bursting with geometric confinements are commonplace in practice. Specifically, bubbles burst in a shallow liquid layer in many realistic configurations, but the effect of such a shallow liquid layer remains poorly understood. For example, during raindrop impact on a porous medium such as soil, gas bubbles may be trapped inside the impacting droplet which can spread and infiltrate into the soil to form a thin layer with a thickness comparable to the bubble size. The subsequent bursting of bubbles in such a liquid layer may produce bioaerosols carrying the pathogens from the soil \cite{joung2015aerosol,joung2017}. In electrolysis, bubbles generated at the electrode can coalescence in proximity of the geometric confinement near the electrode in a similar setup of bubble bursting at the free surface \cite{bashkatov2025electrolyte}. Electrolyte droplets are sprayed into the growing gas bubble and influence the bubble detachment, with potentially relevance to acid mist formation in electrowinning. Furthermore, in annular flow commonly present in engineering applications like refrigeration systems \cite{upot2021scalable} and chemical engineering \cite{cuadros2019characterization}, bubble bursting in a thin film plays an important role, and their contribution to heat and mass transfer has been shown to be substantial \cite{zhou2023role}. In the above scenarios, the bubbles bursts in a shallow liquid layer, yet the influence of such a geometric confinement on the bubble bursting aerosols is barely understood.

In the above studies, a primary mechanism for aerosol generation during bubble bursting is the ejection of a Worthington jet and the subsequent production of jet drops. Due to the importance of the mechanism, the fluid dynamics of such bubble bursting jets has received extensive attention \cite{deike2018dynamics,brasz2018minimum,ganan2021physics,gordillo2023theory}.
However, in contrast to the extensively studied bubble bursting in an unbounded domain, bubble bursting with nearby geometric complications has remained unclear. Previous studies have shown that other nearby bubbles can significantly influence the bursting jet of a bubble, oftentimes resulting in inclined or faster jets \cite{singh2019numerical,lee2022bursting,auregan2026jet}. However, the effect of widely existing solid wall confinements have received limited attention to the best of our knowledge. In this letter, we simulate bubble bursting in a shallow liquid layer above a solid, no-slip wall. We show that the presence of a wall notably decreases the size of the jet drops produced by modifying the flow field around the bubble. Our theoretical analysis shows that the solid wall produces a viscous sticking effect near the bubble bottom, which contributes to the decrease in jet drop size together with the commonly seen wave damping mechanism. Based on those findings, we propose a semi-empirical scaling relating the size of the jet drops to the bubble-wall distance and clarify when the wall effect becomes significant.

\begin{figure*} 
    \centering
    \includegraphics[width=\linewidth]{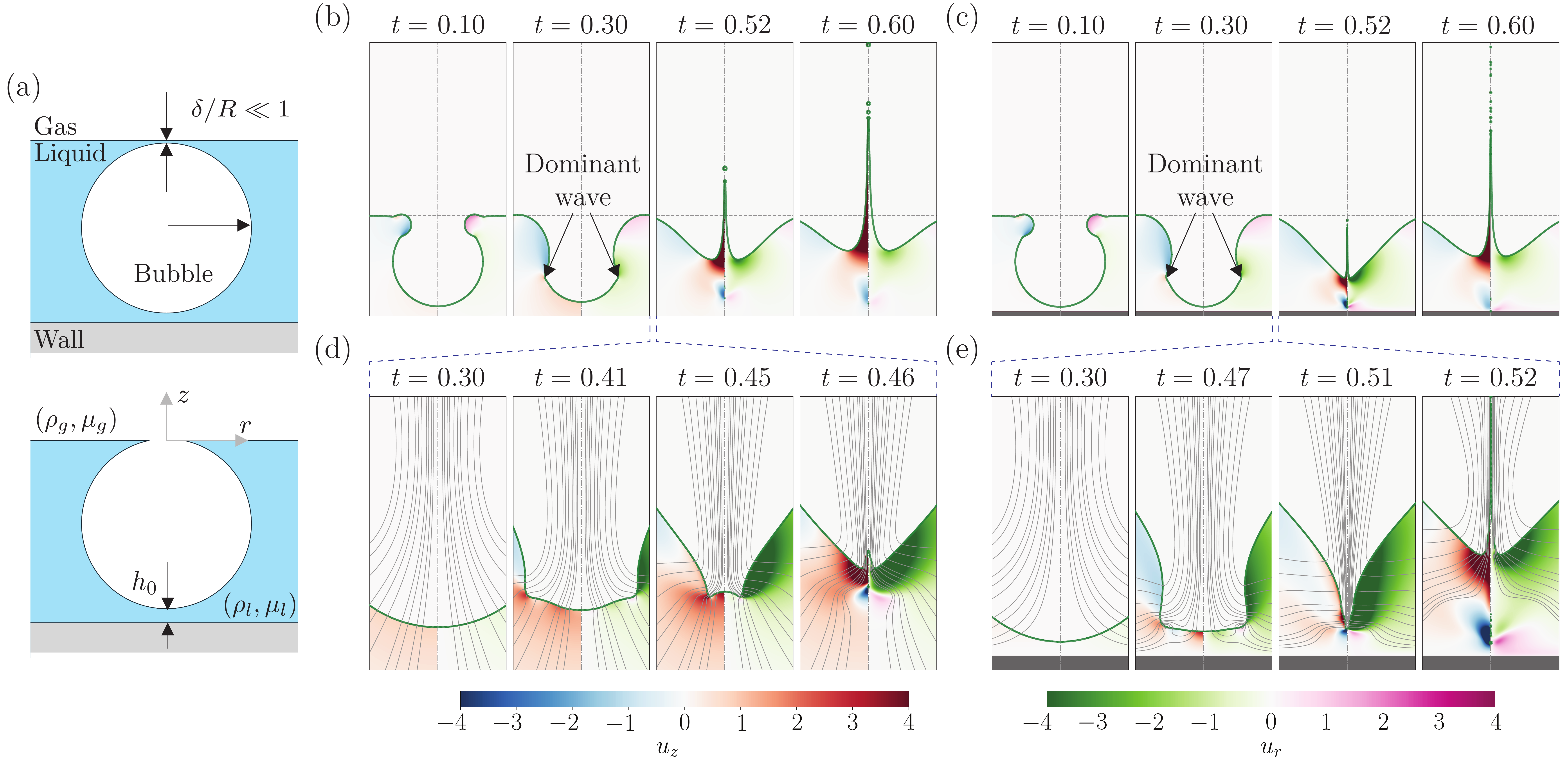}
    \caption{(a) Schematic of the initial simulation setup of bubble bursting in a shallow layer. Top: a bubble resting at the free surface right before the bubble cap film rupture, forming a liquid film with thickness $\delta$. Bottom: bubble cap film rupture initiating the bubble bursting process, used as the initial setup in our simulation. (b-e) Velocity fields during bubble bursting at different initial bubble-wall distances of (b,d) $h_0=2.0$ and (c,e) $h_0=0.1$ at $Oh=0.02$, with (b,c) showing the bursting process of the whole bubbles, and (d,e) showing the zoomed in view of the bubble cavity bottoms. The left and right halves of each individual panel shows axial velocity $u_z$ and radial velocity $u_r$, respectively. The solid body is marked with grey shades. The grey solid lines in (d,e) indicate the streamlines. The grey dashed and dot-dashed lines indicate $z=0$ and $r=0$, respectively.} % \textcolor{red}{can you check simualtion-wise paper in PRL and see what is their first figure? One very clear simulation setup may be needed?}}
    \label{fig:field}
\end{figure*}

Here, we compare the jet drop ejected from bubble bursting at different initial bubble-wall distances via direct numerical simulation. We investigate an axisymmetric bubble collapse in a Newtonian liquid above a solid surface with the open-source Basilisk C program \cite{basilisk}. The configuration is illustrated schematically in Fig. \ref{fig:field}(a) (see Supplemental Materials for details). When a bubble rests at the interface of a shallow liquid layer, film will ultimately rupture and trigger the collapse of the bubble. We initialize the simulations at this onset of cavity collapse. In our simulation, we maintain the Bond number of $Bo = \rho_l g R^2/\gamma = 0.001$ while vary Ohnesorge number $Oh = \mu_l/\sqrt{\rho_l\gamma R}$ between 0.003 and 0.06, where $\rho_l, g, R, \gamma, \mu_l$ represent liquid density, gravitational acceleration, bubble diameter, surface tension and liquid viscosity, respectively. In the following discussions, all parameters are non-dimensionalized by bubble radius $R$, capillary pressure $\gamma/R$, and inertia-capillary velocity $U_c = \sqrt{\gamma/(\rho_l R)}$ (see SM).

Figure \ref{fig:field} (b) and (c) show the bursting dynamics of surface bubbles with the same $Oh = 0.02$ in a deep pool and a shallow layer, respectively. % that a smaller initial bubble-wall distance $h_0$ reduces the size of the ejected drops at fixed $Oh$.
In Fig. \ref{fig:field} (b), the initial bubble-wall distance $h_0 = 2.0$ is chosen to represent a sufficiently deep pool consistent with previous study of Sanjay et al. \cite{sanjay2021}, and its validity will also be discussed later in the current study. Meanwhile, the bubble in Fig. \ref{fig:field} (c) is closer to the bottom wall with a thinner layer as $h_0 = 0.1$. 
% The axial velocity $u_z$ and the radial velocity $u_r$ further reveal the process of bubble bursting 
During bubble bursting, capillary waves propagating down the cavity and drives the collapse of the bubble. In Fig. \ref{fig:field} (b, c), the flow field is further detailed through the axial velocity $u_z$ and the radial velocity $u_r$. When the capillary waves are just initiated during rim retraction ($t=0.10$), the flow near the retracting rim around $z=0$ is directed towards the bulk direction with $u_z<0$ and $u_r>0$. As the distinguishable dominant wave propagates down along the cavity at $t=0.3$, flow behind the dominant wave follows the wave featuring $u_r<0$. Noticeably, a strong upward motion shows at the cavity bottom with $u_z>0$, consistent with the cavity shrinking reported by Krishnan et al. \cite{krishnan2024dynamics}. Finally, the capillary waves focus and form a jet at around $t\approx 0.5$, which eventually breaks into droplets. 
Comparing Fig. \ref{fig:field} (b) and (c), we observe that the radius of the ejected jet drops becomes smaller when the shallow layer exists. As smaller drops remain airborne for longer times, this mechanism of the shallow layer leading to smaller jet drops is of particular interest.

To clarify the reason a closer solid surface results in jet drops of smaller radius, we zoom into the bubble cavity when the jet forms, as shown in Fig. \ref{fig:field} (d,e) where local streamlines are marked. In Fig. \ref{fig:field} (d) where the bubble bursts in a deep pool, the bubble cavity bottom shrinks upward freely before the jet forms with higher $u_z$ and faster interface motion. In comparison, the shrinkage at the bubble cavity bottom is weakened when the bubble bursts in a shallow layer with $h_0 = 0.1$ (Fig. \ref{fig:field} (e)), as if the cavity bottom is sticking to the wall. %With the smaller $u_z$ local wave motions also become more distinct.
The weakened shrinkage with reduced local $u_z$ is a direct result of the solid surface according to the streamline pattern. 
The streamlines cannot penetrate the solid surface and have to bend away as shown in Fig. \ref{fig:field} (e), and therefore the local $u_z$ must be significantly smaller.
The velocity change at the cavity bottom finally alters the morphology of the liquid-gas interface when the jet forms. At the instant of jet formation, the cavity forms a conical shape as capillary waves focus. 
Specifically, if the bubble bursts in a bounded domain, the sticking of the cavity bottom to the wall results in a steeper conical angle at the cavity bottom ($t=0.51$, Fig. \ref{fig:field} (e)) compared to the unbounded case ($t=0.45$, Fig. \ref{fig:field} (d)).
Such a change in the interfacial morphology where the waves focus is recently shown important in the jet production, not only in bubble bursting \cite{gordillo2023theory} but also in related interfacial jetting phenomena such as drop impact \cite{mou2026singular}. Therefore, the wall-induced sticking mechanism that causes the change in the interfacial shape is critical for understanding the bubble bursting jet behaviors. 

%\textit{Dynamics of oil-coated bubble bursting}.-- 
\begin{figure*}[ht]
    \centering
    \includegraphics[width=\linewidth]{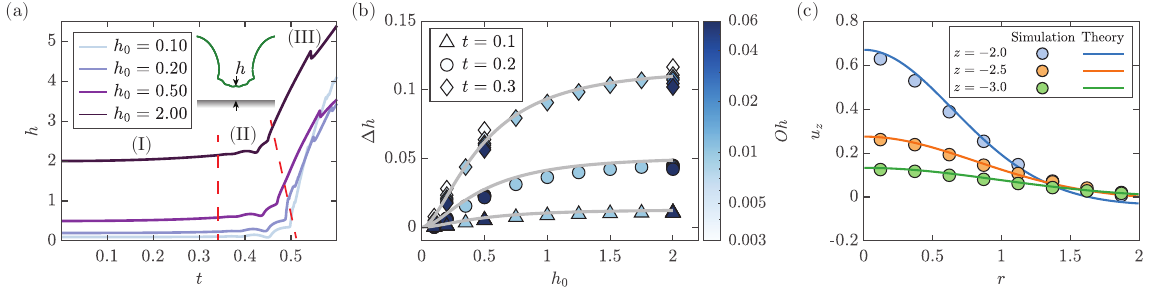}
    \caption{(a) Evolution of distance $h$ between the cavity bottom and the wall as a function of time $t$ for different $h_0$ at $Oh = 0.01$. The regimes (I) cavity shrinking, (II) wave perturbation and (III) jet ejection are separated with red dashed lines. (b) Cavity bottom displacement $\Delta h = h-h_0$ as a function of initial bubble-wall distance. Grey solid curves indicate theoretical prediction from Eq. (\ref{eq:deltaH}). (c) Comparison between simulated and theoretical axial velocity profiles as a function of radial position, where $h_0=2.0$, $Oh=0.02$.}
    \label{fig:thickness}
\end{figure*}

To quantify the sticking effect from the solid surface, we track the evolution of the liquid layer thickness $h$ at cavity bottom where $r=0$, as shown in Fig. \ref{fig:thickness} (a). We divide the evolution of $h$ into three different regimes: (I) cavity shrinking, (II) wave perturbation, and (III) jet ejection. During the first regime of cavity shrinking, $h$ increases monotonously with a slow speed with a small $u_z \lesssim 1$. In the second regime of wave perturbation starting around $t\approx 0.36$, the cavity shrinking is faster in average, while noticeable fluctuation is observed for $h$. The fluctuations appear due to the precursor waves propagating ahead of the dominant wave \cite{krishnan2017scaling}. When the dominant wave arrives and focuses at the cavity bottom near $t\approx 0.5$, the jet is ejected upward. In this third regime after the jet forms, $h$ increases rapidly with the jet speed, and produce spikes when drops pinch off from the jet \cite{sanjay2021}. Both regimes (I) and (II) are influenced by the sticking effect before the jet formation, yet the wave perturbation also influences the cavity shrinking effect in regime (II). Therefore, to isolate the sticking effect before jet ejection, we specifically discuss the regime (I) next.  

To clearly indicate the how much the cavity bottom moves with the influence of sticking effect, we calculate the variation of the bubble-wall distance $\Delta h = h-h_0$ as shown in Fig. \ref{fig:thickness} (b).   
The markers shows the experimental variation of $\Delta h$ as a function of $h_0$ at selected $t$ and $Oh$, and smaller $\Delta h$ indicates stronger sticking effect.
As $h_0$ decreases from 2 to 0.1 at the same $Oh = 0.01$, $\Delta h$ decreases, showing a consistently stronger sticking effect as the initial solid surface is placed closer to the bubble. In contrast, we observe that changes in $Oh$ yield less significant variation of $\Delta h$.
While we change the Ohnesorge number by over one order of magnitude, corresponding to 20 times of change in the dimensional viscosity when the other dimensional parameters are maintained, the resulting variation of $\Delta h$ is still not comparable to that resulted from slight changes in $h_0$.
This weak dependence of $\Delta h$ on $Oh$ can be unexpected in a system of Newtonian liquid, where  the sticking effect is intuitively attributed to viscous effects.
The ineffectiveness of $Oh$, which compares the viscous and inertial effects under the characteristic velocity $U_c$ in bubble bursting, therefore raises a question regarding the mechanism of the cavity bottom sticking and the subsequent jet formation.
% The small effect of $Oh$ leaves us a question. We expect that the sticking effect intuitively originate from the viscous effect of the liquid, yet when we change the Ohnesorge number hence the liquid viscosity by one order of magnitude, $\Delta h$ isn't significantly affected. This therefore leaves a question of why this is the case. 

We consider the interaction between the overall inertia-capillary flow and the viscous effect introduced by the solid surface to elucidate the origin of the sticking effect. 
Specifically, we focus on the flow field beneath the cavity ($z\le -2$), and the case in the absence of the solid surface is investigate first as a reference for further discussions on the cavity sticking in bounded cases. 
For cavity collapse in an unbounded domain prior to jet formation, the characteristic Reynolds number of the overall flow field would be $Re = \rho U_c R/\mu = Oh^{-1} \gg 1$. With the vorticity mainly confined near the interfaces, the flow is mainly irrotational, especially in regime (I) where the cavity bottom has not been perturbed by the capillary waves. This allows us to use the potential flow approximation similar to discussions in previous work \cite{gordillo2019capillary,blanco2021jets}. Here we adopt an ansatz of a quadrupole flow field to describe the flow below the bubble, with the velocity field written as
\begin{equation} \label{eq:quadra}
\left\{
\begin{aligned}
    v_z& = -M(t) \frac{3z(2z^2-3r^2)}{(z^2+r^2)^{7/2}}\\
    v_r &= -M(t) \frac{3r(4z^2-r^2)}{(z^2+r^2)^{7/2}}
\end{aligned}
\right.
.
\end{equation}
Here $M(t)$ is a time-dependent coefficient approximately linear in time as $M(t) = mt$ in regime I, and the subscripts $r,z$ represent the radial and axial directions, respectively (see further discussions in SM). The coordinate origin is defined at the intersection of the undisturbed water interface and the symmetry axis.
To demonstrate that the ansatz sufficiently describes the velocity field beneath the cavity bottom in regime (I), we compares the axial velocity component between the simulation and Eq. (\ref{eq:quadra}) in Fig. \ref{fig:thickness}(c). Good agreement is observed between the theorertical ansatz and the simulation results, indicating that the quadrupole field is a good approximation to the flow beneath bubble cavity. The quadrupole velocity field corresponds to a flow away from the free surface and converging near the bubble bottom. % We further show that in experiments, we have an approximately linear $M(t) = mt$. When $u_zt\ll 1$, we can approximate the rising velocity of the bubble bottom with the velocity $u_{z,b}$ at $z=-2,r=0$, so that $\Delta h \approx \int_0^tu_{z,b}\mathrm{d}t = 3mt^2/16.$

For bubble cavity collapse in the presence of a nearby solid surface such as the case with $h_0 = 0.1$ shown in Fig. \ref{fig:field} (e), the flow below the bubble cavity bottom is much weaker than the rim retraction motion and is confined within the liquid gap between the bubble cavity and the bottom surface. The local Reynolds number $Re_l = \rho \Delta h h_0/(\mu t)$ ranges between $O(0.01 - 1)$, showing that the local dynamics is dominated by viscous effects. Therefore, the method of images for low-Reynolds-number flow can be employed to approximate the local velocity field \cite{blake_note_1971,happel1983low}. The updated field is obtained using the following velocity field given by method of image from a flow with potential (see derivation in SM) as
\begin{equation} \label{eq:mImage}
\left\{
\begin{aligned}
    u_z& = v_z -2z\left({\partial v_z}/{\partial z}\right)\\
    u_r &= v_r +2z\left({\partial v_z}/{\partial r}\right)
\end{aligned}
\right.    .
\end{equation}
The shrinking distance of the cavity bottom is then estimated as $\Delta h \approx \int_0^tu_{z,b}\mathrm{d}t,$ where $u_{z,b}$ is the axial velocity defined at $z=-2,r=0$, the initial position of the cavity bottom. Considering the almost linear $M(t)$, we obtain
\begin{equation} \label{eq:deltaH}
    \Delta h \sim f(h_0)t^2,
\end{equation}
shown as grey curves in Fig. \ref{fig:thickness} (b). 
The coefficient $f(h_0) = 1-(1+h_0)^{-4}-4h_0(1+h_0)^{-5}$ only depends on the initial bubble-wall distance $h_0$ only. The prediction from Eq. (\ref{eq:deltaH}) agrees well with the simulation results in Fig. \ref{fig:thickness} (b). The results demonstrate that for sufficiently small $h_0$ such as $h_0 = 0.1$, the viscous effect remains important for the sticking at the cavity bottom in the whole range of $Oh$. Therefore, the sticking is mainly influenced by $h_0$ instead of $Oh$. Reducing $h_0$ limits the upward motion of the liquid surface at the cavity bottom, further influencing the jet formation process.

Equation (\ref{eq:deltaH}) also implies that the coefficient $f(h_0)$ also quantifies how much shrinking remains when the cavity collapse is influenced by the sticking from the nearby solid surface. For example, when the initial bubble bottom is two radii over the solid surface as is used for the control simulation in the current study and in previous work \cite{sanjay2021}, the normalized $h_0=2$ and the corresponding $f(h_0)$ is $95.5\%$ and close to 1. Therefore, the solid surface barely influences the shrinking of the cavity and the subsequent jet formation. Our results further provides a general criterion that the effect of an ideal solid surface on bubble bursting jet can be neglected when the initial bubble-wall distance exceeds twice radius $h_0>2$. 

\begin{figure} 
    \centering
    \includegraphics[width=0.85\linewidth]{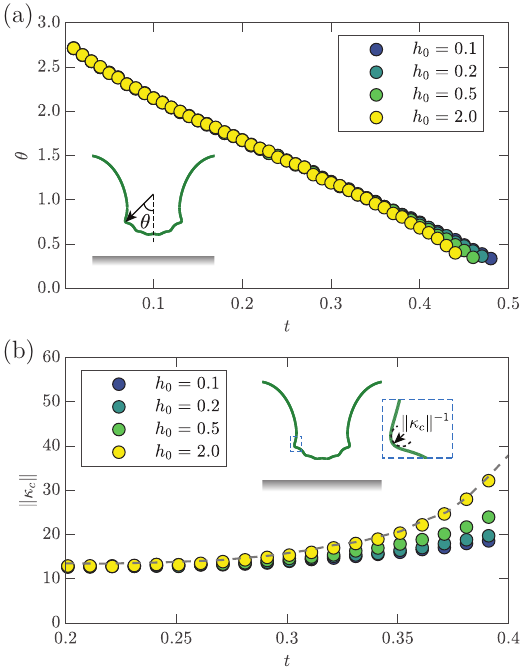}
    \caption{ (a) Wave position angle $\theta$ as a function of time $t$ at different initial $h_0$. (b) Maximum curvature $\|\kappa_c\|$ at the dominant wave trough as a function of time $t$. The grey dashed curve shows the $\|\kappa_c\|$ evolution in a deep pool at the same Ohnesorge number from Sanjay et al. \cite{sanjay2021}. All cases are computed at $Oh=0.01$. }
    \label{fig:wave}
\end{figure}

In addition to the conical angle at jet formation, the wave damping has also been understood to play an important role in influencing the jet radius, with stronger wave damping favoring production of smaller jet drops \cite{deike2018dynamics,gordillo2019capillary}. Therefore, we further examine the propagation of the waves. Figure \ref{fig:wave} (a) shows the temporal evolution of the wave position angle $\theta$. In the cavity shrinking regime, the dominant wave propagating down is far away from the cavity bottom and solid surface and therefore expected to be independent of $h_0$. Consistently, the wave positions at different $h_0$ collapse onto a single curve before $t=0.3$, showing no influence from the liquid layer depth on the wave propagation speed in this regime. 
In the wave perturbation regime when the dominant wave approaches the cavity bottom, the effects from nearby solid surfaces start to emerge. With closer solid surface, the final $\theta-t$ slope decreases, indicating slower final wave speed. 
Meanwhile, Fig. \ref{fig:wave} (b) shows the evolution of
the maximum curvature $\|\kappa_c\|$ at the dominant wave trough as the waves propagate near the cavity bottom, which also represents the strength of the wave \cite{sanjay2021}. 
The wave strength remains almost independent of $h_0$ before $t=0.3$, while its evolution is increasingly dependent on $h_0$ when in the wave perturbation regime. As the shallow layer thickness decreases, the final $\|\kappa_c\|$ also becomes smaller. 
The trend indicates that as the waves approach the cavity bottom, the geometry confinement enhances wave dissipation. The dissipation helps shelter the self-similarity from wave perturbations and contributes to production of thinner jets. 
Based on the above discussion, we summarize the effects of a nearby solid surface on a bubble bursting in a Newtonian liquid. The solid surface barely influences the waves propagating far from the cavity bottom, during which $Oh$ is the main parameter controlling the damping of the capillary waves. 
However, the solid surface results in a sticking effect that cumulatively influences the shrinking of the cavity which finally changes the conical angle. It also enhances the wave disspation as they propagate toward the cavity bottom.
Both effects finally reduce the size of the top jet drops ejected.

\begin{figure} 
    \centering
    \includegraphics[width=\linewidth]{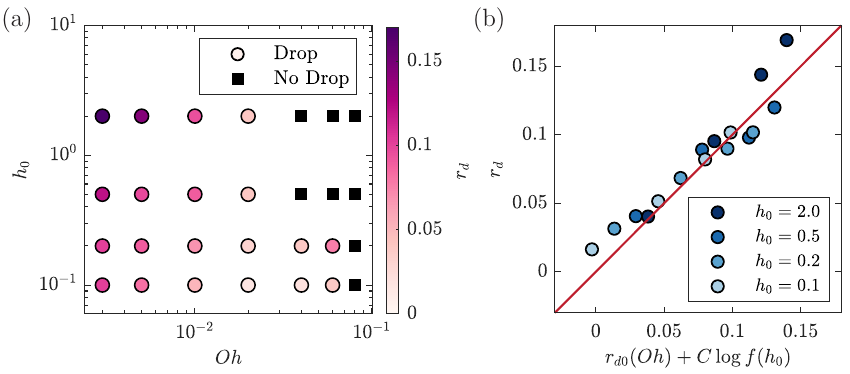}
    \caption{(a) Regime map of jet drop generation in terms of $Oh$ and $h_0$, with the shading of the circle symbols representing the top characteristic jet drop radius $r_d$. (b) Collapse of $r_d$ onto the prediction of $r_{d0}(Oh) + C\log f(h_0)$ from the semi-empirical 
        fitting of Eq. (\ref{eq:rdOh}) indicated by the red line. } %\textcolor{red}{Can we improve the red line? Try a vivid and beautiful way. Right now it is too normal (I will say let's put in the experimental results)}
        %Radius of the top characteristic drop from bubble bursting. (a) Regime map of drop generation in terms of $Oh$ and $h_0$, with the shade of the circle points representing the jet drop radius $r_d$. 
    \label{fig:jetdrop}
\end{figure}

Based on the above findings, we next quantify the influence of the solid wall on jet drop ejection.
Figure \ref{fig:jetdrop} (a) demonstrates a regime map for jet drop ejection regarding Ohnesorge number $Oh$ and the initial bubble-wall distance $h_0$. When $h_0$ decreases below 0.2, the thinner jets from a shallower liquid layer more easily produce jet drops up to $Oh = 0.06$ compared to those from a deep pool. Meanwhile, the top characteristic drop size $r_d$ (see SM) also decreases as $h_0$ decreases. 
We further note that $r_d$ is a non-monotonic function of $Oh$ due to the different roles of viscous effects above and below the critical Ohnesorge number of $0.03$. In the following, we focus on the low Ohnesorge number regime of $Oh<0.03$, where the influences from the viscous effects are less complex, allowing us to individuall inspect how the solid surface affects the jet drop size.  
 % From our results, the shallow layer doesn't influence the wave propagation down the cavity but modifies the cavity shrinking process as well as the wave damping process. When $h_0\rightarrow\infty$ so that $f(h_0)$ approaches 1, the effect of the shallow layer is eliminated, and $r_d$ should converge to the drop size in a deep pool which we denote as $r_{d0}(Oh)$. We show that a semi-empirical fit, $r_d(Oh,h_0)=r_{d0}(Oh)-C\log f(h_0)$, describes the experimental results reasonably well. With the semi-empirical fit, we show that the influence of changes in $Oh$ and $h_0$ can be separated for jet drop radius, where both the increase of $Oh$ and the decrease of $h_0$ reduces the radius of the jet drops. 
 %We also note that $h_0=0.05$ can also disrupt the self-similar collapse though: wave issues at low Ohnesorge number and ? at high Ohnesorge number. (some issues exist here). 

We further propose a semi-empirical scaling law of top jet drop radius from bubble bursting in a shallow liquid layer as a function of $Oh$ and $h_0$. The wave damping is initially mainly governed by $Oh$, while the reduction of $r_d$ due to the solid surface mainly enters through the final modification of the conical angle and the wave damping near the cavity bottom. Therefore, we assume that the deviation in the top jet drop radius $r_d(Oh,h_0)$ from its counterpart in a deep pool $r_{d0}(Oh)$ can be described as a function of the coefficient $f(h_0)$ related to the bubble-wall distance and cavity shrinking. In the limit of $h_0\rightarrow\infty$ so that $f(h_0)\rightarrow 1$, the effect of the shallow layer vanishes, and $r_d$ should converge to $r_{d0}(Oh)$. We show that a consistent fit in Fig. \ref{fig:jetdrop} of
\begin{equation} \label{eq:rdOh}
r_d(Oh,h_0)=r_{d0}(Oh)+C\log f(h_0)
\end{equation}
reasonably well collapses with the numerical results. Here $r_{d0}\propto 1-(Oh/0.0305)^{1/2}$, and $C=0.156$ is a fitting parameter. The fit in Eq. (\ref{eq:rdOh}) we proposed quantitatively describe how the increase of $Oh$ and the decrease of $h_0$ reduce the radius of the jet drops, highlighting the non-negligible influence of a shallow liquid layer on the formation of jet drops.

In summary, we have investigated the influence of a nearby solid wall beneath a bursting bubble on jet drop formation in shallow liquid layers. We show that the presence of the wall significantly reduces the size of the top jet drops, establishing geometric confinement as an important factor governing aerosol generation from bubble bursting. Specifically, the solid wall modifies the local velocity field and induces a viscous sticking effect that suppresses cavity shrinkage, producing a steeper cavity geometry at the instant of jet formation. In addition, the shallow liquid layer enhances capillary-wave damping near the cavity bottom. Together, these effects generate thinner jets and smaller jet drops than bubble bursting in deep liquid pools.  the effects of liquid viscosity and wall proximity and captures the observed variation in jet drop size relatively well when $h_0\ge 0.1$. We note that when the thickness of the shallow liquid layer further decreases, the wall might further influence the self-similar dynamics of wave focusing, and the initial bubble shape may even become constrained by the solid surface. These effects are beyond the scope of the current study and will be considered in future investigation.

%In summary, we have investigated the influence of a nearby solid wall beneath a bursting bubble on the jet drops the bubble produces. We show that the presence of a solid wall can significantly reduce the size of top jet drops, highlighting geometric confinement as an influence factor governing aerosolization from bubble bursting. Specifically, since streamlines cannot penetrate the solid surface, the wall modifies the velocity field, and the viscous sticking from the wall suppresses the cavity shrinking to produce a steeper conical angle when the jet forms. Furthermore, the wave damping is strengthened by the thin liquid layer. The two effects act together, resulting in smaller jet drops when the bubble bursting happens in a shallow liquid layer. We further propose a semi-empirical scaling law that separates the effects of liquid viscosity and wall proximity and captures the observed variation in jet drop size relatively well when $h_0\ge 0.1$. We note that when the thickness of the shallow liquid layer further decreases, the wall might further influence the self-similar dynamics of wave focusing, and the initial bubble shape may even become constrained by the solid surface. These effects are beyond the scope of the current study and will be considered in future investigation.

Our study establishes geometric confinement as an important factor governing jet formation and aerosol generation during bubble bursting. The configuration of bubble bursting over a neighboring solid surface is relevant to realistic scenarios such as bubble coalescence in electrolysis \cite{bashkatov2025electrolyte}, bubble dynamics in annular flow boiling \cite{zhou2023role}, and bubble bursting in droplet impact \cite{maes2025birth}.The physical mechanism identified here has broad implications for predicting aerosol generation as well as heat and mass transfer in these systems. We anticipate that the present framework can be extended to more complex confinement geometries, such as porous or soft confinements, and more broadly, will facilitate the rational control of bubble-bursting aerosols through boundary design.

\bigskip
Z.Y. and J.F. acknowledge partial support by the NSF under grant no. CBET 2323045 and CBET 2426809.  %We thank Prof. Jos\'e M L\'opez-Herrera at for helpful discussions on simulation. %\textcolor{red}{(can we compare with our experiments?? That will be great if it is meaningful to compare. We had some discussions along that)}.

\bibliography{sn-bibliography}
\end{document}